# REALIZATION OF AN ATOMIC MICROWAVE POWER STANDARD

Dave Paulusse, Nelson Rowell and Alain Michaud

Institute for National Measurement Standards, National Research Council of Canada
M-36, 1200 Montreal Road, Ottawa, ON, Canada, K1A-0R6, E-mail: alain.michaud@nrc.ca

## Abstract

We demonstrate the feasibility of a novel microwave power standard based on the electromagnetic interaction with cold atoms. Under the effect of the radiation, the internal state populations will undergo a Rabi oscillation. The measurement of its frequency will allow the determination of the electromagnetic field strength.

## Introduction

The purpose of this paper is to summarize our progress towards using the electromagnetic-atomic interaction as a tool in the field of microwave power measurements [1,2]. The recent development in atomic/laser physics has had major impacts on most fields of metrology. For example the introduction of laser cooling has been a revolution in time and frequency metrology and femtosecond laser science could have a similar impact on dimensional measurements. Furthermore we believe that there are ways of exploiting these new physical methods in microwave measurements as will be discussed.

We are presently building an experiment that will use laser cooled Rubidium atoms subjected to the electromagnetic field. Under the effect of the radiation, the internal state populations will undergo a Rabi oscillation. The measurement of its frequency will allow the determination of the electromagnetic field strength. The following is a short description of the apparatus.

## Novel Atomic Power Standard

The atoms are captured in a standard magneto-optical trap (MOT). The use of cold atoms has several advantages, namely the interaction time and atomic density are increased by many orders of magnitude and the Doppler effect becomes negligible. Simply shutting off the lasers and letting the atoms fall into the interaction region does the measurement.

Before the atoms fall, the magnetic field is ramped to zero, the cooling laser is detuned and the light intensity reduced to further reduce the temperature. Finally a short pulse of resonant light is applied to prepare the internal state of all the atoms. These steps are done in about 20 ms.

When the atoms are in the interaction zone, a pulse of microwave radiation is applied. When the frequency of the field is resonant with the transition (6.8 GHz), only the field amplitude determines the Rabi frequency.

After the atoms have crossed the interaction region, a laser beam probes the population distribution. As this will destroy the atomic coherence, the experiment has to be repeated many times for varying pulse duration times. The population inversion can then be plotted as a function of the time. In principle a continuous wave could replace this pulse. In that case the interaction time would be the time interval between the pumping laser pulse and the detection laser pulse. Alternately the population inversion can be plotted as a function of the field amplitude for a fixed pulse width and is shown in the figure 1.

The microwave field comes from a radiating structure of which the radiation pattern should be known accurately. For that reason we used a simple geometry i.e. the open end of a waveguide terminated by an 'infinite' metallic flange [3].

The waveguide (R-70) attaches directly to the output of a directional coupler that has a power sensor attached to its side arm. Once the experiment is completed the coupler and power sensor can be used directly as a calibrated transfer standard [1,2]. The use of a second coupler would allow the system to stabilize the output power of a source. Some interests have been shown recently for this application [4].

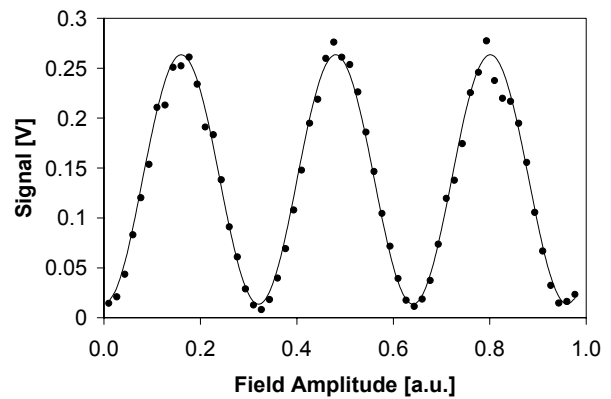

Figure 1 Transition Probability as a function of the electromagnetic field amplitude for a pulse length of 7 ms

## Outline of the presentation

In this paper we report on the progress of this work. We show some schematic describing the experiment and we present our latest experimental results.